\documentclass[prl,superscriptaddress,twocolumn,showpacs,longbibliography]{revtex4-1}
\pdfoutput=1
\usepackage{amsmath, amsthm, amssymb, mathrsfs}
\usepackage[pdfborder={0 0 0}]{hyperref}

\usepackage{graphicx}
\usepackage{amsmath}
\usepackage{amssymb}
\usepackage{epstopdf}
\usepackage{grffile}
\usepackage{array}
\usepackage{tikz}
\usepackage[utf8]{inputenc}

\newcommand{\dd}{\mathrm{d}}
\renewcommand{\vec}[1]{\boldsymbol{#1}}
\newcommand{\up}{\uparrow}
\newcommand{\down}{\downarrow}

\begin{document}

\title{Fulde-Ferrell-Larkin-Ovchinnikov pairing as leading instability on the square lattice}

\author{Jan Gukelberger}
\email[Corresponding author: ]{gukelberger@phys.ethz.ch}
\affiliation{Theoretische Physik, ETH Zurich, 8093 Zurich, Switzerland}

\author{Sebastian Lienert}
\affiliation{Theoretische Physik, ETH Zurich, 8093 Zurich, Switzerland}

\author{Evgeny Kozik}
\affiliation{Physics Department, King's College London, Strand, London WC2R 2LS, United Kingdom}

\author{Lode Pollet}
\affiliation{Department of Physics, Arnold Sommerfeld Center for Theoretical Physics and Center for NanoScience, University of Munich, Theresienstrasse 37, 80333 Munich, Germany}

%\author{Nikolay Prokof'ev}
%\affiliation{Department of Physics, University of Massachusetts, Amherst, MA 01003-4525, USA}
%
%\author{Boris Svistunov}
%\affiliation{Department of Physics, University of Massachusetts, Amherst, MA 01003-4525, USA}

\author{Matthias Troyer}
\affiliation{Theoretische Physik, ETH Zurich, 8093 Zurich, Switzerland}

\date{\today}                                           % Activate to display a given date or no date

\begin{abstract}
We study attractively interacting spin-$\frac{1}{2}$ fermions on the square lattice subject to a spin population imbalance.
Using unbiased diagrammatic Monte Carlo simulations we find an extended region in the parameter space where the Fermi liquid is unstable towards formation of Cooper pairs with non-zero center-of-mass momentum, known as the Fulde-Ferrell-Larkin-Ovchinnikov (FFLO) state.
In contrast to earlier mean-field and quasi-classical studies we provide quantitative and well-controlled predictions on the existence and location of the relevant Fermi-liquid instabilities.
The highest temperature where the FFLO instability can be observed is about half of the superfluid transition temperature in the unpolarized system.
\end{abstract}

% 37.10.Jk 	Atoms in optical lattices
% 67.85.-d 	Ultracold gases, trapped gases
% 71.10.Hf 	Non-Fermi-liquid ground states, electron phase diagrams and phase transitions in model systems
% 74. 	Superconductivity
% 74.20.-z 	Theories and models of superconducting state
% 74.20.Mn 	Nonconventional mechanisms
% 74.20.Rp 	Pairing symmetries (other than s-wave)
% 74.25.Dw 	Superconductivity phase diagrams

\maketitle

Fifty years after the initial prediction by Fulde, Ferrell, Larkin, and Ovchinnikov (FFLO) \cite{larkin1964iss,fulde1964sss}, superconducting phases with spontaneously broken translational invariance are still at the center of interest in such diverse fields as solid state physics, cold atomic gases, nuclear physics, and dense quark matter in neutron stars \cite{matsuda2007fsh,zwicknagel2010bti,Radzihovsky2010,casalbouni2004isc,alford2008csd}.
While the underlying mechanism is generic enough to apply to any partially polarized Fermi system, it has proven surprisingly difficult to unambiguously observe such phases in nature.
Recently, however, experimental evidence has been mounting for their existence in heavy fermion compounds \cite{bianchi2003pfs,radovan2003mes,koutroulakis2010fec} and layered organic materials \cite{beyer2012aef,lortz2007cef,bergk2011mte,wright2011zpt,Mayaffre2014}.
On the other hand, experiments with ultracold atoms, which are among the cleanest imbalanced Fermi systems without the need for a magnetic field, so far failed to demonstrate inhomogeneous superfluidity \cite{zwierlein2006fsi,partridge2006pps} --- although there is some evidence for such a phase in one dimension (1D)  \cite{liao2010sof} --- possibly due to small extent of the parameter region where an FFLO phase may exist in three dimensions (3D) and difficulties in reaching sufficiently low temperatures \cite{Radzihovsky2010}.

On the theoretical side, results on the existence and nature of FFLO phases based on well-controlled microscopic theories are scarce, with the exception of 1D systems, where exact analytical and numerical studies are possible \cite{Buzdin1983,Machida1984,Buzdin1987,orso2007afg,yang2001iss,feiguin2007psp,Batrouni2008}, and where finite-momentum pairing is a generic feature of the spin-imbalanced phase diagram.
In higher dimensions, most studies are based on effective field theories in the neighborhood of critical points or resort to quasi-classical or mean-field approximations.
%For 3D Fermi gases, the ground state phase diagram obtained from the mean-field theory %, containing a homogeneous superfluid at moderate polarization and a paramagnetic normal state at large polarization, separated by a large patch of phase separation,
%\cite{Sheehy2006} has been corroborated by fixed-node diffusion quantum Monte Carlo calculations \cite{Pilati2008}; whether the FFLO phase exists in a small sliver of the phase diagram is not resolved yet. {\color{red} [if the ground state phase diagram has been corroborated then why is there still an open question? What do you want to say?]}
For 3D Fermi gases, many features of the mean-field phase diagram \cite{Sheehy2006} have been corroborated by fixed-node diffusion quantum Monte Carlo calculations \cite{Pilati2008}; whether the FFLO phase does exist in a small sliver of the phase diagram, as predicted by the mean-field theory, is however still subject to debate.

The FFLO state is expected \cite{matsuda2007fsh,zwicknagel2010bti} to occupy a larger parameter region in two dimensions (2D), and lattice effects may further increase its stability \cite{Koponen2007,Loh2010}. Correspondingly, mean-field calculations \cite{chiesa2013pas,baarsma2015lpt} and real-space dynamical mean-field theory (DMFT) for fermions in anisotropic optical lattices find a stable and extended spatially modulated superfluid \cite{Kim2011,Kim2012,Heikkinen2013}.
%, so optical lattice experiments are prime candidates for a detailed study of these unconventional states in a clean and well-controlled environment.
However, such approximations are particularly questionable in 2D. The only numerically exact study to date is a determinantal quantum Monte Carlo simulation of the attractive Hubbard model \cite{Wolak2012}, showing a finite-momentum peak in the pair-momentum distribution in large parts of the polarization--temperature phase diagram.
Unfortunately, this study is severely limited by the negative sign problem and could not reach low enough temperatures to establish phase coherence of the pairs. Therefore, the question of whether or not an FFLO phase with (quasi-)long-range order can emerge in a given microscopic model remains open.

\begin{figure}
    \center
    \includegraphics[width=\columnwidth]{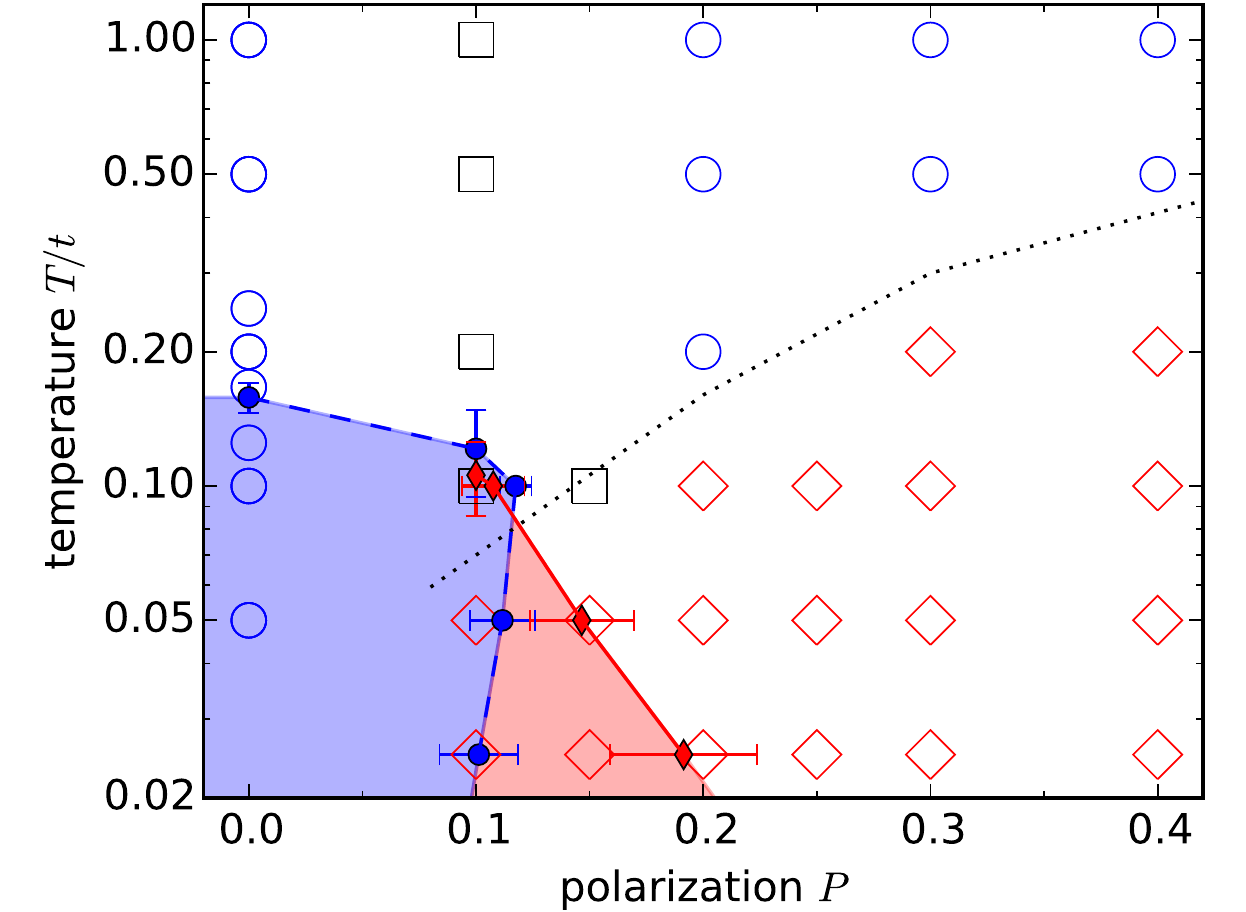} %{./figs/U-4_n0.5_2d-phasediag}
    \caption{(all figures: color online). Phase diagram for $U/t=-4$ at quarter filling: The white region is a Fermi liquid. In the blue shaded region, the Fermi liquid is unstable towards conventional ($\vec{Q}=0$) pairing. In the red shaded region there is an exclusive FFLO instability with finite pair momentum $\vec{Q}_*$.
    Open symbols indicate whether zero- (blue circles) or finite-momentum pairing (red diamonds) is dominant (black squares: no significant difference). The black dotted line separates the two regimes. All lines are guides to the eye.
    }
    \label{fig:phasediag}
\end{figure}

In this Letter we employ the unbiased diagrammatic Monte Carlo (DiagMC) method \cite{prokofev2007bdm,VanHoucke2010,kozik2010} to identify superfluid instabilities of spin-imbalanced fermions on a 2D lattice in a controlled way at lower temperatures than hitherto accessible.
Our main result is that, for attractive interactions of the order of half the bandwidth, there is an extended region in the temperature--polarization plane, indicated by red shading in Fig.~\ref{fig:phasediag}, where the Fermi liquid is unstable exclusively towards FFLO superfluidity.
Specifically, we simulate the Hubbard model on a square lattice
\begin{align}
H = &
-t \sum_{\langle i,j \rangle, \sigma} \left( c_{i \sigma}^\dag c_{j \sigma} + h.c. \right)
+ U \sum_i n_{i \uparrow} n_{i \downarrow} %- \mu \sum_{i,\sigma} n_{i,\sigma}
\label{eq:hamiltonian}
\end{align}
with nearest-neighbor hopping amplitude $t=1$ setting the scale of energy and on-site attraction $U < 0$. Here $c_{i \sigma}^\dag$  and $c_{i \sigma}^{\,}$ are fermionic creation and annihilation operators  with spin $\sigma=\uparrow,\downarrow$, and $n_{i \sigma}=c_{i \sigma}^\dag c_{i \sigma}$.
Spin imbalance is quantified by the polarization
%\begin{align}
%    P = \frac{ \langle n_{i \down} - n_{i \up} \rangle }{ \langle n_{i \down} + n_{i \up} \rangle } ,
%\end{align}
$P = \langle n_{i \down} - n_{i \up} \rangle / \langle n_{i \down} + n_{i \up} \rangle$,
such that $P=1$ corresponds to a fully polarized system.
In the following, we present results for $U=-4$ at quarter filling $n= \langle n_{i \uparrow} + n_{i \downarrow} \rangle = 0.5$. % and comment on other fillings at the end.

Our DiagMC algorithm \cite{VanHoucke2010,kozik2010} stochastically samples many-body Feynman diagrams (built on the bare Green's function) for the self-energy and the two-particle-irreducible pairing vertex directly in the thermodynamic limit. 
Due to the diagrammatic sign problem, in practice a cutoff $N_*$ on the maximum addressed diagram order is introduced and independence of the results on $N_*$ is checked by varying the cutoff. 
We identify continuous phase transitions to ordered phases by monitoring the divergence of the corresponding susceptibilities on approach to the phase boundary from the normal phase. 
According to the Bethe-Salpeter equation, the susceptibility
%\begin{align} \label{eq:bse}
$    \chi_c = \chi^{(0)}_c + \chi^{(0)}_c \Gamma_c \chi_c = \chi^{(0)}_c / \left[1 - \chi^{(0)}_c \Gamma_c \right]$
%\end{align}
in a given channel $c$ (with, e.g., zero or non-zero center-of-mass momentum) diverges when the largest eigenvalue of the kernel $\chi^{(0)}_c \Gamma_c$ reaches unity.
Here, $\chi^{(0)}_c$ denotes the product of two one-particle propagators and $\Gamma_c$ the two-particle-irreducible pairing vertex, see Refs.~\cite{gukelberger2014pss,gukelberger2015diss} for details. 
In the present case, we are primarily interested in pairing of $\up$- with $\down$-particles, which gives rise to both the conventional BCS and the FFLO phases, and hence concentrate on the superconducting channels with total spin projection $S_z=0$ first. We refer to these as ``singlet'' channels \footnote{
    As spin imbalance breaks spin rotation symmetry, there is no distinction between singlet and triplet sectors in the usual sense, but the $S_z=0$ triplet and singlet sectors merge into one $S_z=0$ sector and the $S_z=\pm 1$ sectors become inequivalent. We here refer to pairing between different or identical spins (i.e.\ the $S_z=0$ sector and those with $|S_z|=1$) as singlet or triplet pairing, respectively.
} and compute their pairing eigenvalues $\lambda_{\vec{Q}}$ for different pair momenta $\vec{Q}$.

It is possible that the transition to an FFLO state is actually first-order due to appearance of solid-type order. In this case, the transition temperature extracted from the Bethe-Salpeter equation would correspond to a lower bound. However, the FFLO transition on the 2D lattice is generally believed to be continuous, at least in the neighborhood of the temperature where the FFLO instability first emerges \cite{burkhardt1994fsl,buzdin1997ggt,mora2004nfp,chiesa2013pas,baarsma2015lpt}.

\begin{figure}
    \center
    \includegraphics[width=\columnwidth]{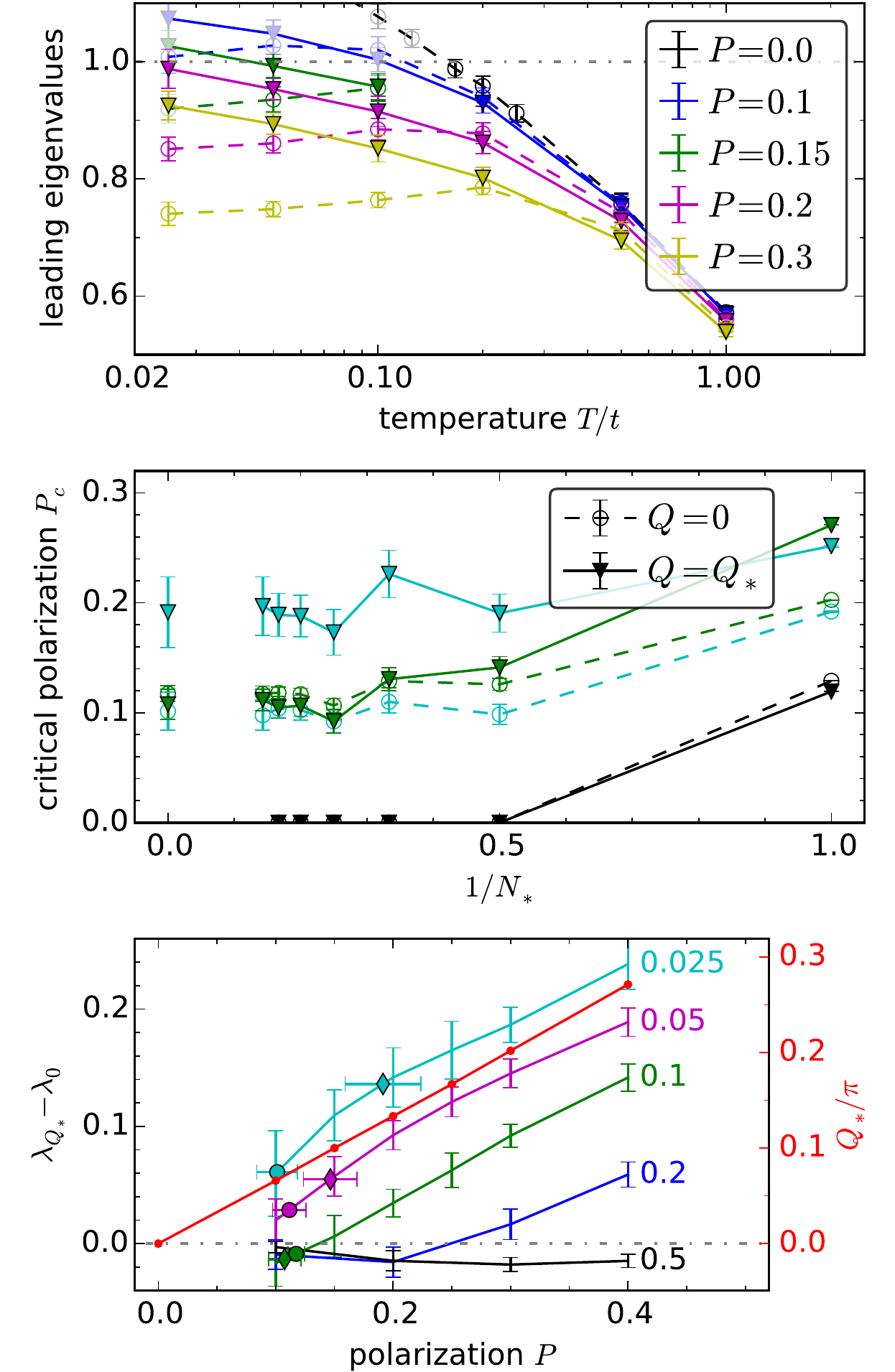} %{./figs/U-4_n0.5_2d-eigvals}
    \caption{\textit{(top)} Temperature dependence of the leading pairing eigenvalues for zero momentum (open symbols/dashed lines) and finite momentum $\vec{Q}_*$ (filled symbols/solid lines) for different polarizations $P$. %The points in lighter shades lie inside the superfluid phase, where our diagrammatic approach is inadequate; we use them as means of extrapolation of the leading order below the transition.
    \textit{(center)} Estimates of the critical polarization for the two channels from varying diagram order cutoff $N_*$ for temperatures $T=0.5$ (black), $T=0.1$ (green), and $T=0.025$ (cyan). The left-most data points show our extrapolations $N_* \to \infty$. % determining the phase boundaries in Fig.~\ref{fig:phasediag}.
    \textit{(bottom)} Difference between FFLO and conventional pairing eigenvalues for varying polarization. The corresponding temperatures are indicated to the right of the curves. 
    Circles and diamonds on the curves indicate the critical polarization where zero- and finite-momentum eigenvalues respectively cross unity. Also shown is the pair momentum magnitude $Q_*$ (red dots).
    }
    \label{fig:eigvals}
\end{figure}

In the following we compare the pairing eigenvalues at $Q=0$ and at a non-zero candidate momentum $\vec{Q}_*$ (to be defined below). 
Studying their temperature dependence, shown in the top panel of Fig.~\ref{fig:eigvals}, we find that a non-zero polarization strongly suppresses the singlet superfluid instabilities as soon as the temperature is low enough to resolve the mismatch between the Fermi surfaces (FSs) of the minority and majority species:
While the transition temperature in the unpolarized system is roughly $T_c = 0.15$, a moderate polarization of $P=0.2$ may only lead to a transition (in the FFLO channel) at the lower end of the considered temperature range, $T_c \lesssim 0.025$.
At larger polarizations $P \gtrsim 0.3$ all the singlet eigenvalues seem to saturate below unity, indicating the absence of a transition in the considered channels at any temperature.

Comparing eigenvalues for zero and non-zero pair momentum, one may differentiate three regimes:
At high temperatures the FSs are so blurred that the two channels are essentially degenerate.
In the region where the effects of the FS mismatch first become noticeable, there is a small advantage of  zero-momentum pairing. % \footnote{The effect is on the order of our error bars, but appears very consistently over different simulations and also on the level of our mean-field analysis (see below).}.
Here a configuration where all parts of one FS are close to the other FS, even if the two never intersect, is apparently more favorable than the alternative with some matching parts and others that are very far apart.
At an even lower temperature, finally, the zero-momentum eigenvalue starts decreasing whereas the finite momentum one continues growing, although with a decreasing rate.
Depending on polarization, the following scenarios are generically realised when the system is cooled down:
(a) For small polarization, the $\vec{Q}=0$ eigenvalue grows to unity before it is overtaken by the $\vec{Q}_*$ eigenvalue.
(b) For larger polarization, the FFLO eigenvalue will reach unity first.
(c) For even larger polarization, all singlet eigenvalues saturate below unity, i.e. the Fermi liquid phase remains stable until triplet pairing emerges at exponentially low temperatures (see below).
%In other words, either of the instabilities may develop first, or the Fermi liquid phase may remain stable until triplet pairing emerges at exponentially low temperatures (see below).

By tracking the growth of the pairing eigenvalues with decreasing polarization at fixed temperature $T$, we find the critical polarization $P_c(T)$ for superfluidity.
Its extrapolation in the diagram-order cutoff, the uncertainty of which is indicated by horizontal error bars on the phase boundaries of Fig.~\ref{fig:phasediag}, is shown in the central panel of Fig.~\ref{fig:eigvals}.
The bottom panel of  Fig.~\ref{fig:eigvals}   shows the difference between non-zero- and zero-momentum pairing eigenvalues.
A positive (negative) difference corresponds to dominant FFLO (conventional) pairing fluctuations in the Fermi liquid phase, indicated by red diamonds (blue circles) in the phase diagram.
Non-zero-momentum pairing fluctuations are dominant at large polarization and low temperature, which is in accord with the large region found in %the DetQMC study of 
Ref.~\cite{Wolak2012} where the pair momentum distribution function is peaked at finite momenta. % but there is no superfluidity % (somewhat confusingly called ``FFLO phase'' in that work).
For temperatures $T \lesssim 0.05$, the difference is positive at the critical polarization $P_c(T)$ implying that the FFLO instability is reached before the conventional superfluid one in this region of the phase diagram \footnote{ A polarized superfluid phase with $\vec{Q}=0$ may also have unconventional properties like the Sarma phase, but this is unlikely for the moderate values of interaction and temperature discussed here \cite{sarma1963iue,Dao2008}.}.
We cannot reliably compute the extent of the FFLO phase because our diagrammatic approach is not valid inside an ordered phase, but we can estimate the region of onset of conventional order in the absence of an FFLO phase by extrapolating the growth of the corresponding pairing eigenvalue from the Fermi liquid phase.
As means of extrapolation we use the finite-order eigenvalue estimates inside the superfluid phase, drawn in lighter shades in Fig.~\ref{fig:eigvals}.

%: If a hypothetical Fermi liquid at a given point in the phase diagram is stable with respect to $Q=0$ pairing, but unstable towards FFLO pairing, the conventional superfluid will not replace the FFLO superfluid.
%In fact, the truncated diagrammatic series, as computed by DiagMC, provides a very natural extrapolation of the Fermi liquid properties, even on the ordered side of a phase transition where the infinite series must diverge:
%n the neighborhood of the phase transition, the pairing instability only becomes manifest in the summation over an infinite number of ladder-type (sub-)diagrams, as done, e.g., by solving the Bethe-Salpeter equation

\begin{figure}
	\center
	\includegraphics[width=\columnwidth]{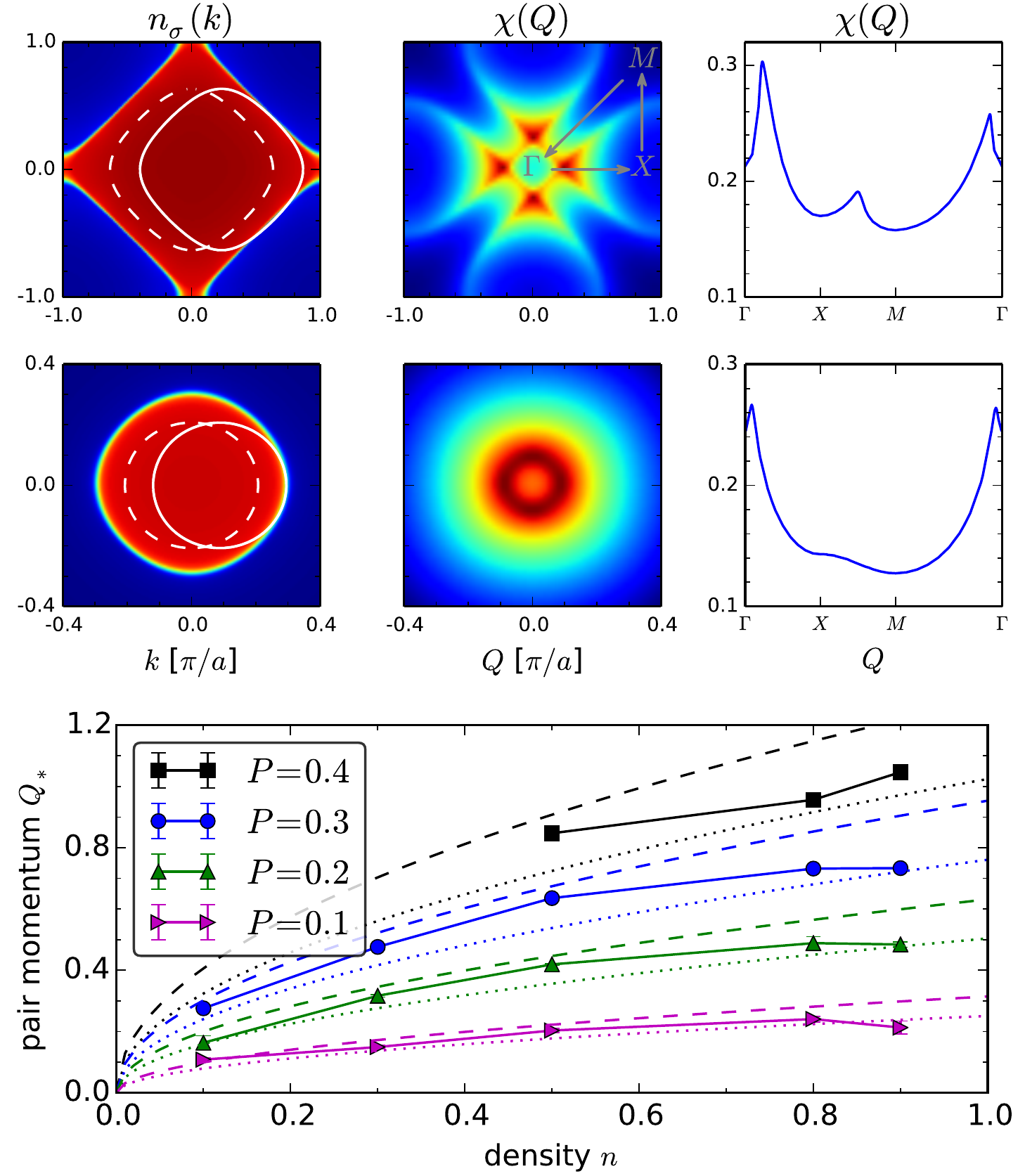}
	\caption{Influence of density shown for $n=0.8$ \textit{(top row)} and $n=0.1$ \textit{(center row)} with polarization $P=0.3$ and temperature $T=0.025$. Left panels show majority spin momentum distribution (colors, from blue=unoccupied to red=occupied) and minority FS (dashed contour), as well as the latter shifted by the optimal pair momentum $\vec{Q_*}$ (solid contour). The other panels illustrate the dependence of the one-particle propagator product $\chi(\vec{Q})$ on the pair momentum $\vec{Q}$. %Brillouin zone plots for $n=0.1$ are zoomed to the central region $k_{x,y}\in[-0.4 \pi,0.4 \pi]$.
	\textit{(bottom)} Dependence of the optimal pair momentum $Q_*$, extracted from the one-particle propagator product $\chi(\vec{Q})$, on density $n$ and polarization $P$. Dotted lines indicate the weak-coupling form for an isotropic dispersion, dashed lines for a square-shaped Fermi surface.}
\label{fig:chiqdep}
\end{figure}

In our study, the optimal pair momentum of the FFLO state $\vec{Q}_*$ could only be determined approximately because
%, within DiagMC, the irreducible vertex needs to be recomputed for every candidate momentum $\vec{Q}$, such that an optimization of the pairing eigenvalue $\lambda_{\vec{Q}}$ over arbitrary pair momenta would be too costly.
an optimization of the pairing eigenvalue $\lambda_{\vec{Q}}$ over $\vec{Q}$ would be too costly within DiagMC.
To this end we replace the irreducible vertex in the Bethe-Salpeter equation by the bare interaction $U$, such that the approximate pairing eigenvalue $ \tilde{\lambda}_{\vec{Q}} = -U \chi(\vec{Q})$ only depends on $\vec{Q}$ via the product of single-particle propagators
\begin{align} \label{eq:lambdaqbcs}
    \chi(\vec{Q}) &= T \sum_n \int \frac{\dd^2 k}{4 \pi^2} \, G_\down(\vec{k},i \omega_n) G_\up(\vec{Q-k},-i \omega_n) ,
\end{align}
which can easily be evaluated numerically. The optimal pair momentum is always found to lie on the coordinate axes of the Brillouin zone.
This is most easily understood close to half filling (top row of Fig.~\ref{fig:chiqdep}), where the FSs are almost squares.
Then, a pair momentum of the form $\vec{Q}_*=(Q_*,0)$ (and those related by point-group symmetry) can connect two sides of the minority FS to the corresponding majority FS patches, whereas, say, a diagonal pair momentum could only connect one side of each FS.
For dilute systems (center row of Fig.~\ref{fig:chiqdep}), the FSs are almost isotropic such that the majority and minority FS can at best touch in one tangential point. The difference between pair momenta with the same magnitude is rather small, but for finite filling we always find a slight preference for pair momenta along the lattice axes.
The bottom panel of Fig.~\ref{fig:chiqdep} plots this pair momentum $Q_*$ found by numerical optimization for different site fillings and polarizations.
In general, there is no closed expression for $Q_*$, but one can consider two limiting cases:
(a) For circular FSs the respective Fermi momenta are
$k_F^\sigma = \sqrt{4 \pi n_\sigma}$,
so the $\up$ and $\down$ FSs are connected by
$Q_* = k_F^\down - k_F^\up = \sqrt{2 \pi n} (\sqrt{1+P} - \sqrt{1-P})$.
(b) For square-shaped FSs, whose corners lie on the coordinate axes at
$k_F^\sigma = \sqrt{2 \pi^2 n_\sigma}$,
the optimal pair momentum is
$Q_* = \sqrt{\pi^2 n} (\sqrt{1+P} - \sqrt{1-P})$.
These estimates are indicated by dotted and dashed lines, respectively, in the bottom panel of Fig.~\ref{fig:chiqdep}.
The data obtained by numerical optimization lie quite consistently between the two extreme estimates. 
%\footnote{
%    Note that the pair momentum $Q_* = \pi n P$, associated with the ``commensurate'' LO state found in the mean field study of Ref.~\cite{chiesa2013pas} close to half filling and for small polarization, agrees with our estimate for square-shaped FSs to leading order in $P$ and $(1-n)$.
%}.
While the approximation $\tilde{\lambda}_{\vec{Q}}$ will in general strongly overestimate the pairing eigenvalue due to the neglect of correlation effects in the vertex, the extracted pair momentum $Q_*$ is expected to be accurate because the momentum dependence of the vertex is typically much weaker than that of the propagators.
Strictly speaking, we cannot exclude the (unlikely) existence of a stronger instability at a different momentum. 
This means that our phase diagram is conservative in the sense that the region of the FFLO phase might become only larger if additional pair momenta are relevant.

\begin{figure}
	\center
	\includegraphics[width=\columnwidth]{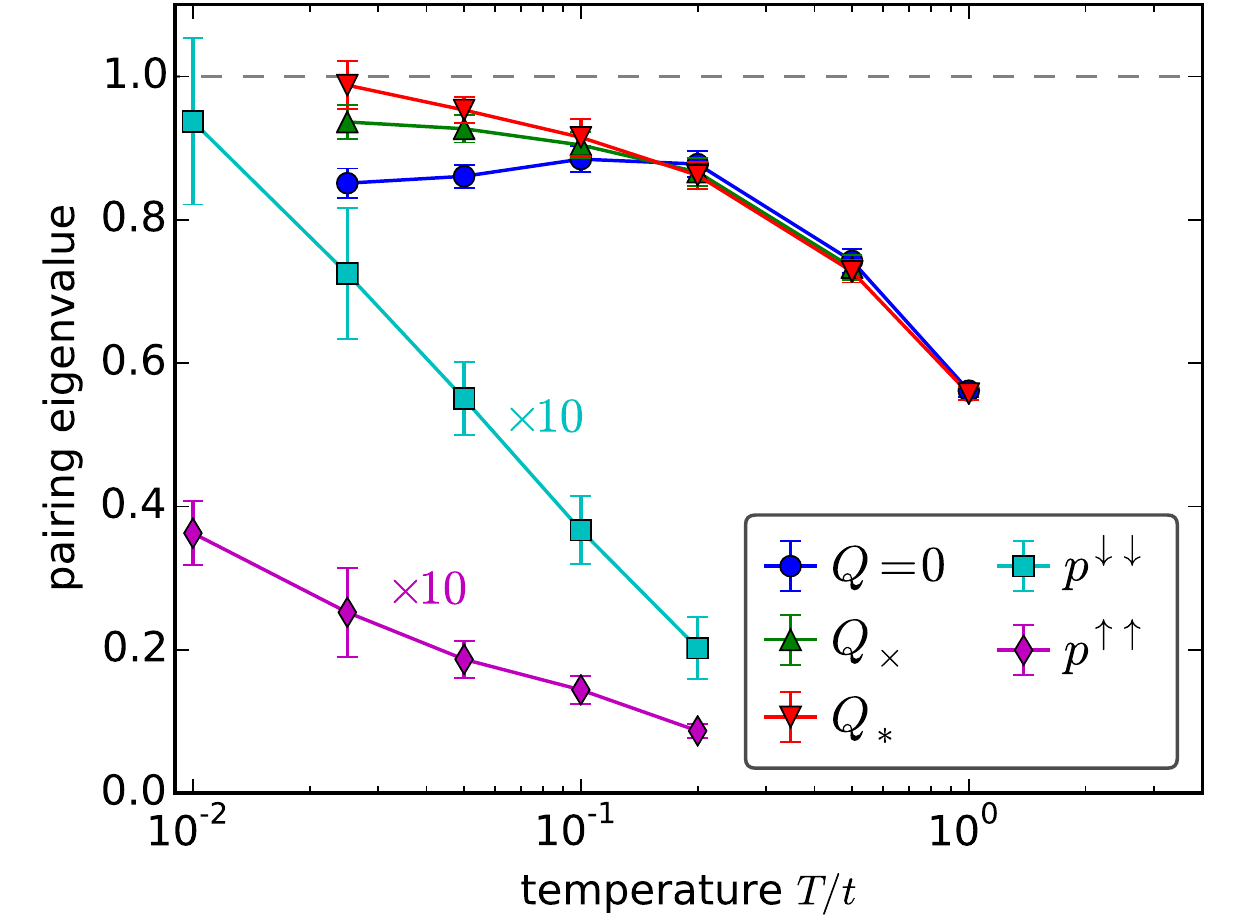} %{./figs/U-4_n0.5_P0.2-alleigvals_vs_T-paper}
	\caption{Pairing eigenvalues for different channels at quarter filling and polarization $P=0.2$. Shown are the singlet pairing eigenvalues with pair momenta $\vec{Q}=0$, $\vec{Q}_*$ (i.e. on the coordinate axis) and $\vec{Q}_\times$ (on the BZ diagonal) as well as the $p$-wave triplet eigenvalues for zero-momentum pairing of the majority ($p^{\down\down}$) and minority ($p^{\up\up}$) species, respectively. The latter have been multiplied by a factor of ten.}
\label{fig:alleigvals}
\end{figure}

At strong polarization and at weaker interaction, all singlet-pairing eigenvalues saturate below unity.
Here, triplet pairing, which is not susceptible to the FS mismatch, will emerge at (exponentially) low temperatures \cite{Konh_Luttinger_theorem} due to an effective interaction between identical particles mediated by the other species, just as in the case of a spin-dependent hopping anisotropy~\cite{gukelberger2014pss}.
In principle, either the majority or the minority species may have the dominant instability.
In second-order perturbation theory at quarter filling, the majority species always reach the superfluid transition first, independent of the polarization.
We have confirmed this with DiagMC calculations for $P=0.2$ (Fig.~\ref{fig:alleigvals}) and $P=0.4$ (not shown).
In Fig.~\ref{fig:alleigvals} we compare the eigenvalues in five different channels:
Among the singlet-pairing eigenvalues, the pair momentum $\vec{Q}_*$ dominates at low temperatures, whereas the conventional $Q=0$ channel saturates below $T \lesssim 0.2$. 
A further candidate, $\vec{Q}_\times$, which is the optimal pair momentum on the BZ diagonal within the approximation \eqref{eq:lambdaqbcs}, is always subdominant to $\vec{Q}_*$. 
The triplet eigenvalues, on the other hand, are by an order of magnitude smaller at the temperatures considered here, but exhibit the logarithmically-diverging growth with decreasing temperature that is standard for a weak-coupling Cooper instability.
As in the weak-coupling calculation, pairing between majority particles clearly dominates over minority pairing.

%\section{Conclusions}

Phase diagrams at densities $n=0.8$ and $0.9$ look very similar to the quarter filled case, indicating that the FFLO instability is not very sensitive to particle density.
For nearly half-filled bands, density-wave instabilities may however become relevant due to nesting in the particle-hole channel; the full phase diagram in the vicinity of half filling is therefore left for further studies.
We have not systematically studied other interactions, but expect $U/t=-4$ to be close to the optimal case for observation of FFLO order: At smaller $|U|$, transition temperatures and the width of the FFLO instability will decrease quickly \cite{paiva2010f2o}, whereas in the strong-coupling regime %$|U| \gg 4t$ 
a (coherent or incoherent) mixture of tightly bound pairs and unpaired excess particles may be more stable than an FFLO phase.
Note that a well-known particle-hole transformation relates the attractive and repulsive Hubbard models to each other \cite{Moreo2007,ho2009qsh}.
Under this transformation, the magnetization $n P$ assumes the role of doping $x=n-1$ and vice versa, whereas the FFLO instability translates into an instability to a striped phase. % in the (doped and spin-imbalanced) repulsive Hubbard model.

Some questions concerning the extent and character of the FFLO state cannot be answered definitively by our study since we cannot enter the broken-symmetry phase. This concerns in particular the type of order (single-$Q$ vs.\ multi-$Q$) and possible transitions between different ordered phases.
We have not detected any hints of phase separation in the polarization vs.\ magnetic field curves, but we cannot conclusively rule this scenario out --- even though previous studies on 2D lattices generally find direct and continuous transitions to the FFLO phase. %\cite{burkhardt1994fsl,buzdin1997ggt,mora2004nfp,chiesa2013pas,baarsma2015lpt}.

In summary, we have presented the first well-controlled numerical evidence for the presence of a Fermi liquid instability towards FFLO order in the spin-imbalanced phase diagram of attractively interacting fermions on a 2D lattice.
For moderate on-site interaction $U/t = -4$, the instability is present in an extended region of the temperature--polarization plane. % for a wide range of lattice fillings.
The largest temperatures where this instability is observable are roughly by a factor of two smaller than the Kosterlitz-Thouless transition temperature in the corresponding spin-balanced system, similar to DMFT results for anisotropic optical lattices \cite{Heikkinen2013}.
%We point out that our phase diagram is conservative in the sense that the actual boundary between the normal and FFLO phases may move to slightly larger polarizations if the transition was first-order and the FFLO-uniform superfluid transition will likely not occur at the place where the conventional Fermi-liquid instability first emerges but rather at smaller polarizations, where uniform and FFLO pairing tendencies are of comparable strength.
At large polarization there does not seem to be any singlet superfluid order and triplet pairing is found at exponentially low temperatures.
Our quantitative predictions may be checked by experiments with cold atoms in optical lattices, where the breaking of translational symmetry can be observed by in-situ imaging \cite{Haller2015,Parsons2015,Cheuk2015} and the pair-momentum distribution by time-of-flight measurements \cite{Regal2004,Zwierlein2004,Perali2005,Altman2005} or in noise correlations \cite{altman2004pms,Luscher2008}.

\begin{acknowledgments}
We acknowledge numerous enlightening discussions with N.\ Prokof'ev and B.\ Svistunov and thank them for helpful comments on the manuscript.
We used the ALPS libraries for simulations and data evaluation \cite{alps20,alps13}.
Simulations were performed on the M\"onch and Brutus clusters of ETH Zurich. This work was supported by FP7/ERC Starting Grant No. 306897, ERC Advanced Grant SIMCOFE and by the Simons Collaboration on the Many Electron Problem. 
\end{acknowledgments}

\bibliography{refs}{}

\end{document}